\pdfoutput=1

\documentclass{article}%
\usepackage{graphicx}
\usepackage{amsmath}
\usepackage{amsfonts}
\usepackage{amssymb}%
\providecommand{\U}[1]{\protect\rule{.1in}{.1in}}

\begin{document}

\author{Antony Valentini\\Augustus College}

\begin{center}
{\LARGE De Broglie-Bohm Pilot-Wave Theory: Many Worlds in Denial?}

\bigskip

Antony Valentini

\bigskip

\textit{Centre de Physique Th\'{e}orique, Campus de Luminy,}

\textit{Case 907, 13288 Marseille cedex 9, France}

and

\textit{Theoretical Physics Group, Blackett Laboratory, Imperial College
London, Prince Consort Road, London SW7 2AZ, United Kingdom.\footnote{Present
address.}}

email: a.valentini@imperial.ac.uk

\bigskip
\end{center}

We reply to claims (by Deutsch, Zeh, Brown and Wallace) that the pilot-wave
theory of de Broglie and Bohm is really a many-worlds theory with a
superfluous configuration appended to one of the worlds. Assuming that
pilot-wave theory does contain an ontological pilot wave (a complex-valued
field in configuration space), we show that such claims arise from not
interpreting pilot-wave theory on its own terms. Specifically, the theory has
its own (`subquantum') theory of measurement, and in general describes a
`nonequilibrium' state that violates the Born rule. Furthermore, in realistic
models of the classical limit, one does not obtain localised pieces of an
ontological pilot wave following alternative macroscopic trajectories: from a
de Broglie-Bohm viewpoint, alternative trajectories are merely mathematical
and not ontological. Thus, from the perspective of pilot-wave theory itself,
many worlds are an illusion. It is further argued that, even leaving
pilot-wave theory aside, the theory of many worlds is rooted in the
intrinsically unlikely assumption that quantum measurements should be modelled
on classical measurements, and is therefore unlikely to be true.

\bigskip

\bigskip

1 Introduction

2 Ontology versus Mathematics

3 Pilot-Wave Theory on its Own Terms

4 Some Versions of the Claim

5 `Microscopic' Many Worlds?

6 `Macroscopic' Many Worlds?

7 Further Remarks

8 Counter-Claim: A General Argument Against Many Worlds

9 Conclusion

\bigskip

\bigskip

To appear in: \textit{Everett and his Critics}, eds. S. W. Saunders \textit{et
al}. (Oxford University Press, 2009).

\bigskip

\bigskip

\bigskip

\bigskip

\bigskip

\bigskip

\bigskip

\bigskip

\bigskip

\bigskip

\bigskip

\bigskip

\bigskip\bigskip

\bigskip

\section{Introduction}

It used to be widely believed that the pilot-wave theory of de Broglie (1928)
and Bohm (1952a,b) had been ruled out by experiments demonstrating violations
of Bell's inequality. Such misunderstandings have largely been overcome, and
in recent times the theory has come to be widely accepted by physicists as an
alternative (and explicitly nonlocal) formulation of quantum theory. Even so,
some workers claim that pilot-wave theory is not really a physically distinct
formulation of quantum theory, that instead it is actually a theory of
Everettian many worlds. The principal aim of this paper is to refute that
claim. We shall also end with a counter-claim, to the effect that Everett's
theory of many worlds is unlikely to be true, as it is rooted in an
intrinsically unlikely assumption about measurement.

Pilot-wave theory is a first-order, nonclassical theory of dynamics, grounded
in configuration space. It was first proposed by de Broglie, at the 1927
Solvay conference (Bacciagaluppi and Valentini 2009). From de Broglie's
dynamics, together with an assumption about initial conditions, it is possible
to derive the full phenomenology of quantum theory, as was first shown by Bohm
in 1952.

In pilot-wave dynamics, a closed system with configuration $q$ has a wave
function $\Psi(q,t)$ --- a complex-valued field on configuration space obeying
the Schr\"{o}dinger equation $i\partial\Psi/\partial t=\hat{H}\Psi$. The
system has an actual configuration $q(t)$ evolving in time, with a velocity
$\dot{q}\equiv dq/dt$ determined by the gradient $\nabla S$ of the phase $S$
of $\Psi$ (for systems with standard Hamiltonians $\hat{H}$).\footnote{More
generally, $\dot{q}=j/|\Psi|^{2}$ where $j$ is the current associated with the
Schr\"{o}dinger equation (Struyve and Valentini 2008).} In principle, the
configuration $q$ includes all those things that we normally call `systems'
(particles, atoms, fields) as well as pieces of equipment, recording devices,
experimenters, the environment, and so on.

Let us explicitly write down the dynamical equations for the case of a
nonrelativistic many-body system, as they were given by de Broglie (1928). For
$N$ spinless particles with positions $\mathbf{x}_{i}(t)$ and masses $m_{i}$
($i=1,2,....,N$), in an external potential $V$, the total configuration
$q=(\mathbf{x}_{1},\mathbf{x}_{2},....,\mathbf{x}_{N})$ evolves in accordance
with the de Broglie guidance equation%
\begin{equation}
m_{i}\frac{d\mathbf{x}_{i}}{dt}=\mathbf{\nabla}_{i}S \label{geqn}%
\end{equation}
(where $\hslash=1$ and $\Psi=\left\vert \Psi\right\vert e^{iS}$), while the
`pilot wave' $\Psi$ (as it was originally called by de Broglie) satisfies the
Schr\"{o}dinger equation%
\begin{equation}
i\frac{\partial\Psi}{\partial t}=\sum_{i=1}^{N}-\frac{1}{2m_{i}}\nabla_{i}%
^{2}\Psi+V\Psi\ . \label{Seqn}%
\end{equation}
Mathematically, these two equations define de Broglie's dynamics --- just as,
for example, Maxwell's equations and the Lorentz force law may be said to
define classical electrodynamics.

The theory was revived by Bohm in 1952, though in a pseudo-Newtonian form.
Bohm regarded the equation%
\begin{equation}
m_{i}\frac{d^{2}\mathbf{x}_{i}}{dt^{2}}=-\mathbf{\nabla}_{i}(V+Q) \label{neqn}%
\end{equation}
as the true law of motion, with a `quantum potential'%
\[
Q\equiv-\sum_{i=1}^{N}\frac{1}{2m_{i}}\frac{\nabla_{i}^{2}\left\vert
\Psi\right\vert }{\left\vert \Psi\right\vert }%
\]
acting on the particles. (Taking the time derivative of (\ref{geqn}) and using
(\ref{Seqn}) yields (\ref{neqn}).) On Bohm's view, (\ref{geqn}) is not a law
of motion but a condition $\mathbf{p}_{i}=\mathbf{\nabla}_{i}S$ on the initial
momenta --- a condition that happens to be preserved in time by (\ref{neqn}),
and which could in principle be relaxed (leading to corrections to quantum
theory) (Bohm 1952a, pp. 170--71). One should therefore distinguish between de
Broglie's first-order dynamics of 1927, defined by (\ref{geqn}) and
(\ref{Seqn}), and Bohm's second-order dynamics of 1952, defined by
(\ref{neqn}) and (\ref{Seqn}). In particular, Bohm's rewriting of de Broglie's
theory had the unfortunate effect of making it seem much more like classical
physics than it really was. De Broglie's original intention had been to depart
from classical dynamics at a fundamental level, and indeed the resulting
theory is highly non-Newtonian. As we shall see, it is crucial to avoid making
classical assumptions when interpreting the theory.

Over an ensemble of quantum experiments, beginning at time $t=0$ with the same
initial wave function $\Psi(q,0)$ and with a Born-rule or `quantum
equilibrium' distribution%
\begin{equation}
P(q,0)=\left\vert \Psi(q,0)\right\vert ^{2} \label{Br}%
\end{equation}
of initial configurations $q(0)$, it follows from de Broglie's dynamics that
the distribution of final outcomes is given by the usual Born rule (Bohm
1952a,b). On the other hand, for an ensemble with a `quantum nonequilibrium'
distribution%
\begin{equation}
P(q,0)\neq\left\vert \Psi(q,0)\right\vert ^{2}\ , \label{notBr}%
\end{equation}
in general one obtains a distribution of final outcomes that
\textit{disagrees} with quantum theory (for as long as $P$ has not yet relaxed
to $|\Psi|^{2}$, see below) (Valentini 1991a,b, 1992, 1996, 2001, 2002, 2004a;
Pearle and Valentini 2006).

The initial distribution (\ref{Br}) was assumed by both de Broglie and Bohm,
and subsequently most workers have regarded it as one of the axioms of the
theory. As we shall see, this is a serious mistake that has led to numerous
misunderstandings, and is partially responsible for the erroneous claim that
pilot-wave theory is really a theory of many worlds.

We shall not attempt to provide an overall assessment of the relative merits
of de Broglie-Bohm pilot-wave theory and Everettian many-worlds theory.
Instead, here we focus on evaluating the following claim --- hereafter
referred to as `the Claim' --- which has more or less appeared in several
places in the literature (Deutsch 1996, Zeh 1999, Brown and Wallace 2005)
(author's paraphrase):

\begin{itemize}
\item Claim: \textit{If one takes pilot-wave theory seriously as a possible
theory of the world, and if one thinks about it properly and carefully, one
ought to see that it really contains many worlds --- with a superfluous
configuration }$q$\textit{ appended to one of those worlds}.
\end{itemize}
Were the Claim correct, one could reasonably add a corollary to the
effect that one should then drop the superfluous configuration $q$, and arrive
at (some form of) many-worlds theory.

Deutsch's way of expressing the Claim has inspired the title of this paper
(Deutsch 1996, p. 225):

\begin{quote}
In short, pilot-wave theories are parallel-universes theories in a state of
chronic denial.
\end{quote}

We should emphasise that here we shall interpret pilot-wave theory (for a
given closed system) as containing an ontological --- that is, physically real
--- complex-valued field $\Psi(q,t)$ on configuration space, where this field
drives the motion of an actual configuration $q(t)$. The Claim asserts that,
if the theory is regarded in these terms, then proper consideration shows that
$\Psi$ contains many worlds, with $q$ amounting to a superfluous appendage to
one of the worlds. One might try to side-step the Claim by asserting that
$\Psi$ has no ontological status in pilot-wave theory, that it merely provides
a mathematical account of the motion $q(t)$. In this case, one could not even
begin to make the Claim, for the complete ontology would be defined by the
configuration $q$. For all we currently know, this view might turn out to be
true in some future derivation of pilot-wave theory from a deeper theory. But
in pilot-wave theory as we know it today --- the subject of this paper ---
such a view seems implausible and physically unsatisfactory (see below). In
any case, even if only for the sake of argument, let us here assume that the
pilot wave $\Psi$ is ontological, and let us show how the Claim may still be refuted.

It will be helpful first to review the distinction between ontological and
mathematical structure in current physical theory, and then to give a brief
overview of pilot-wave theory interpreted on its own terms.

Generally speaking, theories should be evaluated on their own terms,
\textit{without assumptions that make sense only in rival theories}. We shall
see that, in essence, the Claim in fact arises from not interpreting and
understanding pilot-wave theory on its own terms.

\section{Ontology versus Mathematics}

Physics provides many examples of the distinction between ontological and
mathematical structure. Let us consider three.

(1)\textit{ Classical mechanics}. This may be formulated in terms of a
Hamiltonian trajectory $(q(t),p(t))$ in phase space. For a given individual
system, there is only one real trajectory. The other trajectories,
corresponding to alternative initial conditions $(q(0),p(0))$, have a purely
mathematical existence. Similarly, in the Hamilton-Jacobi formulation, the
Hamilton-Jacobi function $S(q,t)$ is associated with a whole family of
trajectories (with $\dot{q}$ determined by $\nabla S$), only one of which is realised.

(2) \textit{A test particle in an external field}. This provides a
particularly good parallel with pilot-wave theory. A charged test particle,
placed in an external electromagnetic field $\mathbf{E}(\mathbf{x},t)$,
$\mathbf{B}(\mathbf{x},t)$, will follow a trajectory $\mathbf{x}(t)$. One
would normally say that the field is real, and that the realised particle
trajectory is real; while the alternative particle trajectories (associated
with alternative initial positions $\mathbf{x}(0)$) are not real, even if they
might be said to be contained in the mathematical structure of the
electromagnetic field. Similarly, if a test particle moves along a geodesic in
a background spacetime geometry, one can think of the geometry as ontological,
and the mathematical structure of the geometry contains alternative geodesic
motions --- but again, only one particle trajectory is realised, and the other
geodesics have a purely mathematical existence.

(3) \textit{A classical vibrating string}. Consider a string held fixed at the
endpoints, $x=0$, $L$. (This example will also prove relevant to the quantum
case.) A small vertical displacement $\psi(x,t)$ obeys the partial
differential equation%
\[
\frac{\partial^{2}\psi}{\partial t^{2}}=\frac{\partial^{2}\psi}{\partial
x^{2}}%
\]
(setting the wave speed $c=1$). This is conveniently solved using the standard
methods of linear functional analysis. One may define a Hilbert space of
functions $\psi$, with a Hermitian operator $\hat{\Omega}=-\partial
^{2}/\partial x^{2}$ acting thereon. Solutions of the wave equation may then
be expanded in terms of a complete set of eigenfunctions $\phi_{m}%
(x)=\sqrt{2/L}\sin\left(  m\pi x/L\right)  $, where $\hat{\Omega}\phi
_{m}=\omega_{m}^{2}\phi_{m}$ with $\omega_{m}^{2}=\left(  m\pi/L\right)  ^{2}$
($m=1,2,3,....$). Assuming for simplicity that $\dot{\psi}(x,0)=0$, we have
the general solution%
\[
\psi(x,t)=\sum_{m=1}^{\infty}c_{m}\phi_{m}(x)\cos\omega_{m}t\ \ \ \ \ \left(
c_{m}\equiv\int_{0}^{L}dx\ \phi_{m}(x)\psi(x,0)\right)
\]
or (in bra-ket vector notation)%
\[
\left\vert \psi(t)\right\rangle =\sum_{m=1}^{\infty}\left\vert m\right\rangle
\langle m\left\vert \psi(0)\right\rangle \cos\omega_{m}t
\]
(where $\hat{\Omega}\left\vert m\right\rangle =\omega_{m}^{2}\left\vert
m\right\rangle $). Any solution may be written as a superposition of
oscillating `modes'. Even so, the true ontology consists essentially of the
total displacement $\psi(x,t)$ of the string (perhaps also including its
velocity and energy). Individual modes in the sum would not normally be
regarded as physically real. One would certainly not assert that $\psi$ is
composed of an ontological multiplicity of strings, with each string vibrating
in a single mode. Instead one would say that, in general, the eigenfunctions
and eigenvalues have a mathematical significance only.

All this is not to say that the question of ontology in physical theories is
trivial or always obvious. On the contrary, it is not always self-evident as
to whether mathematical objects in our physical theories should be assigned
ontological status or not. For example, classical electrodynamics may be
viewed in terms of a field theory (with an ontological electromagnetic field),
or in terms of direct action-at-a-distance between charges (where the
electromagnetic field is merely an auxiliary field, if it appears at all).
Most physicists today prefer the first view, probably because the field seems
to contain a lot of independent and contingent structure (see below).

The question to be addressed here is: in the pilot-wave theory of de Broglie
and Bohm, if one regards the pilot wave $\Psi$ as ontological (which seems the
most natural view at present), does this amount to an ontology of many worlds?

\section{Pilot-Wave Theory on its Own Terms}

In the author's view, pilot-wave theory continues to be widely misinterpreted
and misrepresented, even by some of its keenest supporters. Here, for
illustration, we confine ourselves to de Broglie's original dynamics for a
system of nonrelativistic (and spinless) particles, defined by (\ref{geqn})
and (\ref{Seqn}).

\begin{center}
\textit{Basic History}
\end{center}

Let us begin by setting the historical record straight,\footnote{For a
detailed account, see chapter 2 of Bacciagaluppi and Valentini (2009).} as
historical arguments sometimes play a role in evaluations of pilot-wave theory.

Pilot-wave dynamics was constructed by de Broglie in the period 1923--27. His
motivations were grounded in experiment. He wished to explain the quantisation
of atomic energy levels and the interference or diffraction of single photons.
To this end, he proposed a unification of the physics of particles with the
physics of waves. De Broglie argued that Newton's first law of motion had to
be abandoned, because a particle diffracted by a screen does not touch the
screen and yet does not move in a straight line. During 1923--24, de Broglie
then proposed a new, non-Newtonian form of dynamics in which the
\textit{velocity} of a particle is determined by the phase of a guiding wave.
As a theoretical guide, de Broglie sought to unify the classical variational
principles of Maupertuis ($\delta\int m\mathbf{v}\cdot d\mathbf{x}=0$, for a
particle with velocity $\mathbf{v}$) and of Fermat ($\delta\int dS=0$, for a
wave with phase $S$). The result was the guidance equation (\ref{geqn}) (at
first applied to a single particle and later generalised), which de Broglie
regarded as the basis of a new form of dynamics.

At the end of a rather complicated development in the period 1925--27
(including a crucial contribution by Schr\"{o}dinger, who found the correct
wave equation for de Broglie's waves), de Broglie proposed the many-body
dynamics defined by (\ref{geqn}) and (\ref{Seqn}). De Broglie regarded his
theory as provisional, much as Newton regarded his own theory of gravity as
provisional. And de Broglie regarded the observation of electron diffraction,
by Davisson and Germer in 1927, as a vindication of his prediction (first made
in 1923), and as clear evidence for his new (first-order) dynamics of particle motion.

Clearly, de Broglie's construction of pilot-wave dynamics was motivated by
experimental puzzles and had its own internal logic. Note in particular that
de Broglie did not construct his theory to `solve the measurement problem',
nor did he construct it to provide a (deterministic or realistic) `completion
of quantum theory': for in 1923, there was no measurement problem and there
was no quantum theory.

Getting the history right is important, for its own sake and also because some
criticisms of pilot-wave theory are based on a mistaken appraisal of history.
For example, Deutsch (1986, pp. 102--103) has said the following about the theory:

\begin{quote}
.... to append to the quantum formalism an additional structure .... solely
for the purpose of interpretation, is I think a very dangerous thing to do in
physics. These structures are being introduced solely to solve the
interpretational problems, without any physical motivation. .... the chances
of a theory which was formulated for such a reason being right are extremely remote.
\end{quote}
But there is no sense in which de Broglie `appended' something to
quantum theory, for quantum theory did not exist yet. And de Broglie had ample
physical motivation, grounded in experimental puzzles and in a compelling
analogy between the principles of Maupertuis and Fermat.

A proper historical account also undermines discussions in which pilot-wave
theory is presented as being motivated by the desire to `solve the measurement
problem'. For example, Brown and Wallace (2005) --- who discuss Bohm's
motivations but ignore de Broglie's --- argue that many-worlds theory provides
a more natural solution to the measurement problem than does pilot-wave
theory. The discussion is framed as if the measurement problem were the prime
motivation for considering pilot-wave theory in the first place. As a matter
of historical fact, this is false.

The widespread misleading historical perspective has been exacerbated by some
workers who present de Broglie's 1927 dynamics as a way to `complete' quantum
theory by adding trajectories to the wave function (D\"{u}rr \textit{et al}.
1992, 1996), an approach that furthers the mistaken impression that the theory
is a belated reformulation of an already-existing theory. Matters are further
confused by some workers who refer to de Broglie's first-order dynamics by the
misnomer `Bohmian mechanics', a term that should properly be applied to Bohm's
second-order dynamics. De Broglie's dynamics pre-dates quantum theory; and it
was given in final form in 1927, not as an after-thought (or reformulation of
quantum theory) in 1952.

We may then leave aside certain spurious objections that are grounded in a
mistaken version of historical events. In the author's view, the proper way to
pose the question addressed in this paper is: \textit{given} de Broglie's
dynamics (as it was in 1927), if we examine it carefully on its own terms,
does it turn out to contain many worlds?

\ 

$\ $

\begin{center}
\textit{Basic Ontology}
\end{center}

As stated in the introduction, we regard the theory as having a dual ontology:
the configuration $q(t)$ together with the pilot wave $\Psi\lbrack q,t]$. We
need to give the relation between this ontology and what we normally think of
as physical reality.

De Broglie constructed the theory as a new dynamics of particles:
specifically, the basic constituents of matter and radiation (as understood at
the time). It is then natural to assume that physical systems, apparatus,
people, and so on, are `built from' the configuration $q$. (In extensions of
the theory, $q$ may of course include configurations of fields, the geometry
of 3-space, strings, or whatever may be thought of as the modern fundamental
constituents. Further, macroscopic systems --- such as experimenters --- will
usually supervene on $q$ under some coarse-graining.) This view has been
explicitly stated in the literature by several workers --- for example Bell
(1987, p. 128), Valentini (1992, p. 26), Holland (1993, pp. 337, 350), and
others --- though perhaps it is not clearly stated in some of the de
Broglie-Bohm literature (as Brown and Wallace (2005) suggest). In any case, we
shall take this to be the correct and natural viewpoint.

That $\Psi$ is also to be regarded as ontological is often not explicitly
stated. A notable exception was Bell (1987, p. 128, original italics):

\begin{quote}
.... the wave is supposed to be just as `real' and `objective' as say the
fields of classical Maxwell theory .... . \textit{No one can understand this
theory until he is willing to think of }$\psi$\textit{ as a real objective
field .... . Even though it propagates not in 3-space but in 3N-space}.
\end{quote}

Could $\Psi$ instead be regarded as `fictitious', that is, as a merely
mathematical field appearing in the law of motion for $q$? As already
mentioned, this does not seem reasonable, at least not for the theory in its
present form, where --- like the electromagnetic field --- $\Psi$ contains a
lot of independent and contingent structure, and is therefore best regarded as
part of the state of the world (Valentini 1992, p. 17; Brown and Wallace 2005,
p. 532).

Valentini (1992, p. 13) considered the possibility that $\Psi$ might merely
provide a convenient mathematical summary of the motion $q(t)$; to this end,
he drew an analogy between $\Psi$ and physical laws such as Maxwell's
equations, which also provide a convenient mathematical summary of the
behaviour of physical systems. On this view, `the world consists purely of the
evolving variables $X(t)$, whose time evolution may be summarised
mathematically by $\Psi$' (\textit{ibid}., p. 13). But Valentini argued
further (p. 17) that such a view did not do justice to the physical
information stored in $\Psi$, and he concluded instead that $\Psi$ was a new
kind of causal agent acting in configuration space (a view that the author
still takes today). The former view, that $\Psi$ is law-like, was adopted by
D\"{u}rr \textit{et al}. (1997).\footnote{`.... the wave function is a
component of physical law rather than of the reality described by the law'
(D\"{u}rr \textit{et al}. 1997, p. 33).} They proposed further that the time
dependence and contingency of $\Psi$ --- properties that argue for it to be
ontological (see Brown and Wallace 2005, p. 532) --- may be illusions, as the
wave function for the whole universe is (so they claim) expected to be static
and unique. However, the present situation in quantum gravity indicates that
solutions for $\Psi$ (satisfying the Wheeler-DeWitt equation and other
constraints) are far from unique, and display the same kind of contingency
(for example in cosmological models) that we are used to for quantum states
elsewhere in physics (Rovelli 2004). Should the universal wave function be
static --- and the notorious `problem of time' in quantum gravity urges
caution here --- this alone is not enough to establish that it should be
law-like: contingency, or under-determination by physical law, is the more
important feature.\footnote{One should also guard against the idea ---
sometimes expressed in this context --- that the existence of `only one
universe' somehow suggests that the universal wave function cannot be
contingent. Equally, in non-Everettian cosmology, there is only one
intergalactic magnetic field, and yet it would be generally agreed that the
precise form of this field is a contingency (not determined by physical law).}
Therefore, current theoretical evidence speaks against the idea. And in any
case, our task here is to consider the theory we have now, not ideas for
theories that we may have in the future: in the present form of pilot-wave
theory, the time-dependence and (especially) the contingency of $\Psi$ makes
it best regarded as ontological.

Note that in 1927 de Broglie regarded $\Psi$ as providing --- as a temporary
measure --- a mathematically convenient and phenomenological summary of
motions generated from a deeper theory, in which particles were singular
regions of 3-space waves (Bacciagaluppi and Valentini 2009, section 2.3.2). De
Broglie hoped the theory would later be derived from something deeper (as
Newton believed of gravitational attraction at a distance). Should this
eventually happen, ontological questions will have to be addressed anew.
Alternatively, perhaps de Broglie's `deeper theory' (the theory of the double
solution) should be regarded merely as a conceptual scaffolding which he used
to arrive at pilot-wave theory, and the scaffolding should now be
forgotten.\footnote{Cf. the role played by the ether in electromagnetism, or
in Newton's thinking about gravitation. For a discussion of this parallel, see
section 2.3.2 of Bacciagaluppi and Valentini (2009).} But in any case, the
theory has come to be regarded as a theory in its own right, and the question
at hand is whether \textit{this} theory contains many worlds or not.

\begin{center}
\textit{Equilibrium and Nonequilibrium}
\end{center}

Many workers take the quantum equilibrium distribution (\ref{Br}) as an axiom,
alongside the laws of motion (\ref{geqn}) and (\ref{Seqn}). It has been argued
at length that this is incorrect and deeply misleading (Valentini 1991a,b,
1992, 1996, 2001, 2002; Valentini and Westman 2005; Pearle and Valentini
2006). A postulate concerning the distribution of initial conditions has no
fundamental status in a theory of dynamics. Instead, quantum equilibrium is to
pilot-wave dynamics as thermal equilibrium is to classical dynamics. In both
cases, equilibrium may be understood as arising from a process of relaxation.
And in both cases, the equilibrium distributions are mere contingencies, not
laws: the underlying theories allow for more general distributions, that
violate quantum physics in the first case and thermal-equilibrium physics in
the second.

Taken on its own terms, then, pilot-wave theory is \textit{not} a mere
alternative formulation of quantum theory. Instead, the theory itself tells us
that quantum physics is a special case of a much wider `nonequilibrium'
physics (with $P\neq|\Psi|^{2}$), which may exist for example in the early
inflationary universe, or for relic particles that decoupled soon after the
big bang, or for particles emitted by black holes (Valentini 2004b, 2007, 2008a,b).

\begin{center}
\textit{True (Subquantum) Measurements}
\end{center}

The wider physics of nonequilibrium has its own theory of measurement ---
`subquantum measurement' (Valentini 1992, 2002; Pearle and Valentini 2006).
This is to be expected, since measurement is theory-laden: given a (perhaps
tentative) theory, one should look to the theory itself to tell us how to
perform correct measurements (cf. section 8).

In pilot-wave theory, an `ideal subquantum measurement' (analogous to the
ideal, non-disturbing measurement familiar from classical physics) enables an
experimenter to measure a de Broglie-Bohm system trajectory without disturbing
the wave function. This is possible if the experimenter possesses an apparatus
whose `pointer' has an arbitrarily narrow nonequilibrium distribution
(Valentini 2002, Pearle and Valentini 2006). Essentially, the system and
apparatus are allowed to interact so weakly that the joint wave function
hardly changes; yet, the displacement of the pointer contains information
about the system configuration, information that is visible if the pointer
distribution is sufficiently narrow. A sequence of such operations allows the
experimenter to determine the system trajectory without disturbing the wave
function, to arbitrary accuracy.

\begin{center}
\textit{Generally False Quantum `Measurements' (Formal Analogues of Classical
Measurements)}
\end{center}

We are currently unable to perform such true measurements, because we are
trapped in a state of quantum equilibrium. Instead, today we generally carry
out procedures that are known as `quantum measurements'. This terminology is
misleading, because such procedures are --- at least according to pilot-wave
theory --- generally \textit{not} correct measurements: they are merely
experiments of a certain kind, designed to respect a formal analogy with
\textit{classical} measurements (cf. Valentini 1996, pp. 50--51).

Thus, in classical physics, to measure a system variable $\omega$ using an
apparatus pointer $y$, Hamilton's equations tell us that we should switch on a
Hamiltonian $H=a\omega p_{y}$ (where $a$ is a coupling constant and $p_{y}$ is
the momentum conjugate to $y$). One obtains trajectories $\omega(t)=\omega
_{0}$ and $y(t)=y_{0}+a\omega_{0}t$. From the displacement $a\omega_{0}t$ of
the pointer, one may infer the value of $\omega_{0}$. An experimental
operation represented by $H=a\omega p_{y}$ then indeed realises a correct
measurement of $\omega$ (according to classical physics). But there is no
reason to expect the same experimental operation to constitute a correct
measurement of $\omega$ for a nonclassical system. Even so, remarkably,
so-called quantum `measurements' are in general designed using classical
measurements as a guide. Specifically, in quantum theory, to measure an
observable $\omega$ using an apparatus pointer $y$, one switches on a
Hamiltonian operator $\hat{H}=a\hat{\omega}\hat{p}_{y}$. The quantum procedure
is obtained, in effect, by `quantising' the classical procedure.

But what does this analogous quantum procedure actually accomplish? According
to pilot-wave theory, it merely generates a branching of the total wave
function, with branches labelled by eigenvalues $\omega_{n}$ of the linear
operator $\hat{\omega}$, and with the total configuration $q(t)$ ending in the
support of one of the (non-overlapping) branches. Thus, for example, if the
system is a particle with position $x$, the initial wave function%
\[
\Psi_{0}(x,y)=\left(  \sum_{n}c_{n}\phi_{n}(x)\right)  g_{0}(y)
\]
(where $\hat{\omega}\phi_{n}=\omega_{n}\phi_{n}$ and $g_{0}$ is the initial
(narrow) pointer wave function) evolves into%
\[
\Psi(x,y,t)=\sum_{n}c_{n}\phi_{n}(x)g_{0}(y-a\omega_{n}t)\ .
\]
The effect of the experiment is simply to create this branching.\footnote{Over
an ensemble, if $x$ and $y$ have an initial distribution $P_{0}%
(x,y)=\left\vert \Psi_{0}(x,y)\right\vert ^{2}$, one of course finds that a
fraction $\left\vert c_{n}\right\vert ^{2}$ of trajectories $q(t)=(x(t),y(t))$
end in the (support of) the $n$th branch $\phi_{n}(x)g_{0}(y-a\omega_{n}t)$.}

From a pilot-wave perspective, the eigenvalues $\omega_{n}$ have no particular
ontological status: we simply have a complex-valued field $\Psi$ on
configuration space, obeying a linear wave equation, whose time evolution may
in some situations be conveniently analysed using the methods of linear
functional analysis (as we saw for the classical vibrating string).

It cannot be sufficiently stressed that, generally speaking, by means of this
procedure one has \textit{not measured anything} (so pilot-wave theory tells
us). In quantum theory, if the pointer is found to occupy the $n$th branch, it
is common to assert that therefore `the observable $\omega$ has the value
$\omega_{n}$'. But in pilot-wave theory, all that has happened is that, at the
end of the experiment, the system trajectory $x(t)$ is guided by the
(effectively) reduced wave function $\phi_{n}(x)$.\footnote{Because the
branches have separated in configuration space, it follows from de Broglie's
equation of motion that the `empty' branches no longer affect the trajectory.}
This does not usually imply that the system has or had some property with
value $\omega_{n}$ (at the end of the experiment or at the beginning), because
in pilot-wave theory there is no general relation between eigenvalues and
ontology.\footnote{For example, the eigenfunction $\phi_{E}(x)\propto
(e^{ipx}+e^{-ipx})$ of the kinetic-energy operator $\hat{p}^{2}/2m$ has
eigenvalue $E=p^{2}/2m\neq0$; and yet, the actual de Broglie-Bohm kinetic
energy vanishes, ${\frac{1}{2}}m\dot{x}^{2}=0$ (since $\partial S/\partial
x=0$). If the system had this initial wave function, and we performed a
so-called `quantum measurement of kinetic energy' using a pointer $y$, then
the initial joint wave function $\phi_{E}(x)g_{0}(y)$ would evolve into
$\phi_{E}(x)g_{0}(y-aEt)$ and the pointer would indicate the value $E$ ---
even though the particle kinetic energy was and would remain equal to zero.
The experiment has not really measured anything.}

Thus, a so-called `ideal quantum measurement of $\omega$' is not a true
measurement (a notable exception being the case $\omega=x$). And in general,
it is usually incorrect to identify eigenvalues with values of real physical
quantities: one must beware of `eigenvalue realism'.

\section{Some Examples of the Claim}

Before evaluating the Claim, let us quote some examples of it from the literature.

First, Deutsch (1996, p. 225) argues that parallel universes are

\begin{quote}
.... a logical consequence of Bohm's `pilot-wave' theory (Bohm [1952]) and its
variants (Bell [1986]). .... The idea is that the `pilot-wave' .... guides
Bohm's single universe along its trajectory. This trajectory occupies one of
the `grooves' in that immensely complicated multidimensional wave function.
The question that pilot-wave theorists must therefore address, and over which
they invariably equivocate, is what are the \textit{unoccupied} grooves? It is
no good saying that they are merely a theoretical construct and do not exist
physically, for they continually jostle both each other and the `occupied'
groove, affecting its trajectory (Tipler [1987], p. 189). .... So the
`unoccupied grooves' must be physically real. Moreover they obey the same laws
of physics as the `occupied groove' that is supposed to be `the' universe. But
that is just another way of saying that they are universes too. .... In short,
pilot-wave theories are parallel-universes theories in a state of chronic denial.
\end{quote}

Zeh (1999, p. 200) puts the matter thus:

\begin{quote}
It is usually overlooked that Bohm's theory contains the \textit{same} `many
worlds' of dynamically separate branches as the Everett interpretation (now
regarded as `empty' wave components), since it is based on precisely the same
(`\textit{absolutely} real') global wave function .... . Only the `occupied'
wave packet itself is thus meaningful, while the assumed classical trajectory
would merely point at it: `This is where \textit{we} are in the quantum world.'
\end{quote}

Similarly, Brown and Wallace (2005, p. 527) write the following:

\begin{quote}
.... the corpuscle's role is minimal indeed: it is in danger of being
relegated to the role of a mere epiphenomenal `pointer', irrelevantly picking
out one of the many branches defined by decoherence, while the real story ---
dynamically and ontologically --- is being told by the unfolding evolution of
those branches. The `empty wavepackets' in the configuration space which the
corpuscles do not point at are none the worse for its absence: they still
contain cells, dust motes, cats, people, wars and the like.
\end{quote}

In the case of Zeh, and of Brown and Wallace, the key assertion is that
pilot-wave theory and many-worlds theory contain the same multitude of
wave-function branches, and that in pilot-wave theory the `empty' branches
nevertheless constitute parallel worlds (which `still contain cells, dust
motes, cats, people, wars and the like').

Deutsch's argument leads to the same assertion --- if one interprets his word
`grooves' to mean what are normally called `branches'. However, Deutsch may in
fact have used `grooves' to mean the set of de Broglie-Bohm trajectories, in
which case his version of the Claim states that pilot-wave theory is really a
theory of `many de Broglie-Bohm worlds'.\footnote{Deutsch cites the rather
confused paper by Tipler (1987), which argues among other things that de
Broglie-Bohm trajectories must affect each other in unphysical ways. Tipler's
critique is mostly aimed at a certain stochastic version of pilot-wave theory.
While it is not really relevant to Deutsch's argument, for completeness we
note that, as regards conventional (deterministic)\ pilot-wave theory,
Tipler's critique stems from an elementary misunderstanding of the role of
probability in the theory.} (This version of the Claim is addressed in section
7.) In any case, in essence Deutsch argues that the unoccupied grooves are
real, and that they `obey the same laws of physics' as the occupied groove,
thereby constituting a `multiverse'.

Today, it is often said that in Everettian quantum theory the notion of
parallel `worlds' or `universes' applies only to the macroscopic worlds
defined (approximately) by decoherence. Formerly, it was common to assert the
existence of many worlds at the microscopic level as well. Without entering
into any controversy that might still remain about this, here for completeness
we shall address the Claim for both `microscopic' and `macroscopic' cases.

\section{`Microscopic' Many Worlds?}

In pilot-wave theory, is there a multiplicity of parallel worlds at the
microscopic level? To see that there is not, let us consider some examples.

(1) \textit{Superposition of eigenvalues}. Let a single particle moving in one
dimension have the wave function $\psi(x,t)\propto e^{-iEt}\left(
e^{ipx}+e^{-ipx}\right)  $, which is a mathematical superposition of two
distinct eigenfunctions of the momentum operator $\hat{p}=-i\partial/\partial
x$. Are there in any sense two particles, with two different momenta $+p$ and
$-p$? Clearly not. While the field $\psi\propto\cos px$ has two Fourier
components $e^{ipx}$ and $e^{-ipx}$, there is only one single-valued field
$\psi$ (as in our example of the classical vibrating string). And a true
(subquantum) measurement of the particle trajectory $x(t)$ would reveal that
the particle is at rest (since $S=-Et$ and $\partial S/\partial x=0$). In a
so-called `quantum measurement of momentum', at the end of the experiment
$x(t)$ is guided by $e^{ipx}$ or $e^{-ipx}$: during the experiment the
particle is accelerated and \textit{acquires} a momentum $+p$ or $-p$, as
could be confirmed by a true subquantum measurement. Any impression that there
may be two particles present arises from a mistaken belief in eigenvalue realism.

(2) \textit{Double-slit experiment}. Let a single particle be fired at a
screen with two slits, where the incident wave function $\psi$ passes through
both slits, leading to an interference pattern on the far side of the screen.
Are there in any sense two particles, one passing through each slit? Again,
clearly not. There is a single-valued field $\psi$ passing through both slits,
and there is one particle trajectory $\mathbf{x}(t)$ in 3-space, passing
through one slit only (as again could be tracked by a true subquantum measurement).

(3) \textit{Superposition of `Ehrenfest' packets for a hydrogen atom}.
Finally, consider a single hydrogen atom, with a centre-of-mass trajectory
$\mathbf{x}(t)$ and with a wave function that is a superposition%
\[
\psi=\frac{1}{\sqrt{2}}\left(  \psi_{1}+\psi_{2}\right)
\]
of two localised and spatially-separated `Ehrenfest' packets $\psi_{1}$ and
$\psi_{2}$. Each packet, with centroid $\left\langle \mathbf{x}\right\rangle
_{1}$ or $\left\langle \mathbf{x}\right\rangle _{2}$, follows an approximately
classical trajectory, and let us suppose that the actual trajectory
$\mathbf{x}(t)$ lies in $\psi_{2}$ only. Is there any sense in which we have
\textit{two} hydrogen atoms? The answer is no, because, once again, a true
subquantum measurement could track the unique atomic trajectory $\mathbf{x}%
(t)$ (without affecting $\psi$).

This last example has a parallel in the macroscopic domain, to be discussed in
the next section. Before proceeding, it will prove useful to consider the
present example further. In particular, one might argue that each packet
$\psi_{1}$ and $\psi_{2}$ \textit{behaves like} a hydrogen atom, under
operations defined by changes in the external potential $V$. Specifically, the
motion of the empty packet $\psi_{1}$ will respond to changes in $V$, in
exactly the same way as will the motion of the occupied packet $\psi_{2}$. One
might then claim that, if one regards each packet as physically real, one may
as well conclude that there really are two hydrogen atoms present. But this
argument fails, because the similarity of behaviour of the two packets holds
only under the said restricted class of operations (that is, modifying the
classical potential $V$). In pilot-wave theory, in principle, other
experimental operations are possible, under which the behaviours of $\psi_{1}$
and $\psi_{2}$ will be quite different.

For example, suppose one first carries out an ideal subquantum measurement,
which shows that the particle is in the packet $\psi_{2}$. One may then carry
out an additional experiment --- say an ordinary quantum experiment, using a
piece of macroscopic apparatus --- designed to find out whether or not a given
packet is occupied. One may \textit{predict} that, in the second experiment,
if the operation is performed on packet $\psi_{1}$ the apparatus pointer will
point to `unoccupied', while if the operation is performed on $\psi_{2}$ the
pointer will point to `occupied'.\footnote{In quantum theory too, of course,
the second experiment will always give different results for the two packets.
But the outcome will be random, making the operational difference between the
packets less clear.} It will then become operationally apparent that $\psi
_{1}$ consists solely of a bundle of the complex-valued $\psi$-field, whose
centroid happens to be \textit{simulating} the approximately classical motion
of a hydrogen atom in an external field (under the said restricted class of operations).

It is of course hardly mysterious that in some circumstances one may have an
ontological but empty $\psi$-packet whose motion approximately traces out the
trajectory of a classical body --- just as, in some circumstances, a localised
classical electromagnetic pulse travelling through an appropriate medium (with
variable refractive index) might trace out a trajectory similar to that of a
moving body. In both cases, it would be clear from other experiments that the
moving pulse is not really a moving body.

\section{`Macroscopic' Many Worlds?}

Let us now ask if there is any sense in which pilot-wave theory contains many
worlds at the `macroscopic' level.

We shall begin with an utterly unrealistic example, involving a superposition
of two `Ehrenfest' packets each (supposedly) representing a classical
macroscopic world. This example has the virtue of illustrating the Claim in
what we believe to be its strongest possible form. We shall see that, even for
this example, the Claim may be straightforwardly refuted, along the lines
given in the last section for the case of the hydrogen atom.

We then turn to a further unrealistic example, involving a superposition of
two delocalised `WKB' packets which, again, are each supposed to represent a
classical macroscopic world. This example has the virtue of showing that, if
one cannot point to some piece of localised `$\Psi$-stuff' following an
alternative classical trajectory, then the Claim simply cannot be formulated.
The lesson learned from this example is then readily applied to realistic
cases with decoherence, for which the wave functions involved are also
generally delocalised, and for which, therefore, the Claim again cannot be formulated.

\begin{center}
\textit{The Claim in a `Strong Form'}
\end{center}

Let us again consider an `Ehrenfest' superposition%
\[
\Psi(q,t)=\frac{1}{\sqrt{2}}\left(  \Psi_{1}(q,t)+\Psi_{2}(q,t)\right)  \ ,
\]
where now the configuration $q$ represents not just a single hydrogen atom but
all the contents of a macroscopic region --- for example, a region including
the Earth, with human experimenters, apparatus, and so on. We shall imagine
that the centroids $\left\langle q\right\rangle _{1}$, $\left\langle
q\right\rangle _{2}$ of the respective packets $\Psi_{1}$, $\Psi_{2}$ follow
approximately classical trajectories, corresponding to alternative histories
of events on Earth. This is of course not at all a realistic formulation of
the classical limit for a complex macroscopic system: wave packets spread, and
they do so particularly rapidly for chaotic systems. But we shall ignore this
for a moment, because the example is nevertheless instructive.

Let us assume that $\Psi$ consists initially of a single narrow packet, and
that the subsequent splitting of $\Psi$ into the (non-overlapping) branches
$\Psi_{1}$, $\Psi_{2}$ occurs as a result of a `quantum measurement' with two
possible outcomes $+1$ and $-1$. (See Fig. 1.) One might imagine that, at
first, the branches $\Psi_{1}$, $\Psi_{2}$ develop a non-overlap with respect
to the apparatus pointer coordinate $y$, which then generates a non-overlap
with respect to other (macroscopic) degrees of freedom --- beginning, perhaps,
with variables in the eye and brain of the experimenter who looks at the
pointer. We may imagine that it had been decided in advance that if the
outcome were $+1$, the experimenter would stay at home; while if the outcome
were $-1$, the experimenter would go on holiday. These alternative histories
for the experimenter are supposed to be described by the trajectories of the
narrow packets $\Psi_{1}$ and $\Psi_{2}$ (whose arguments include all the
relevant variables, constituting the centre-of-mass of the experimenter, his
immediate environment, the plane he may or may not catch, and so on). Let us
assume that the actual de Broglie-Bohm trajectory $q(t)$ ends in the support
of $\Psi_{2}$, as shown in Fig. 1.%

\begin{figure}
[ptb]
\begin{center}
\includegraphics[
natheight=5.729400in,
natwidth=9.375400in,
height=2.041in,
width=3.3217in
]%
{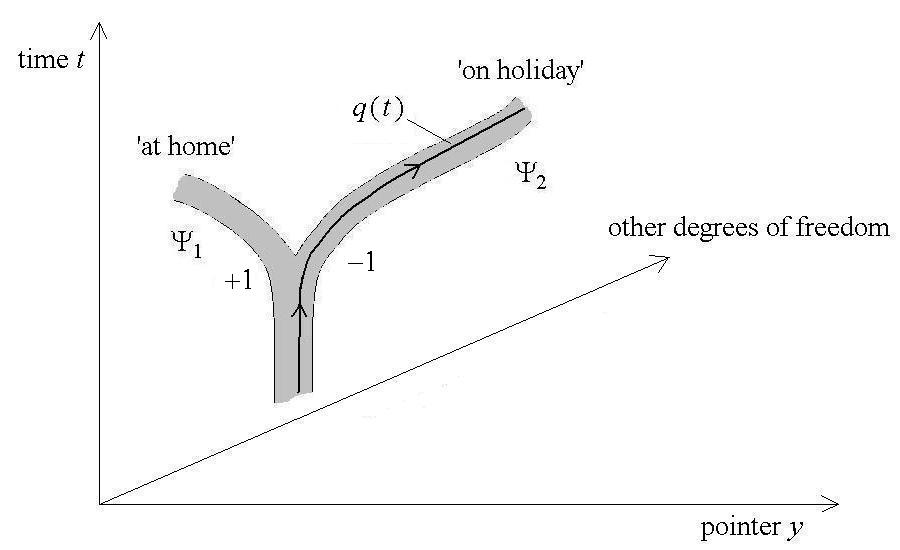}%
\caption{The Claim in a `strong form'.}%
\end{center}
\end{figure}

One could of course extend the example to superpositions of the form
$\Psi=\Psi_{1}+\Psi_{2}+\Psi_{3}+....$, where $\Psi_{1}$, $\Psi_{2}$,
$\Psi_{3}$.... are non-overlapping narrow packets that trace out --- in
configuration space --- approximately classical motions corresponding to
alternative macroscopic histories of the world, with each history containing,
in the words of Brown and Wallace, `cells, dust motes, cats, people, wars and
the like'.

Now, with these completely unrealistic assumptions, the Claim seems to be at
its strongest. For if $\Psi$ is ontological, then in the example of Fig. 1 the
narrow packets $\Psi_{1}$ and $\Psi_{2}$ are both real objects moving along
approximately classical paths in configuration space. There is certainly
\textit{something real} moving along each path. One of the paths has an extra
component too --- the actual configuration $q(t)$ --- but even so the fact
remains that something real is moving along the other path as well.

This situation seems to be the strongest possible realisation of the Claim.
One might say, for example with Brown and Wallace (section 4 above), that
`[t]he `empty wavepackets' in the configuration space which the corpuscles do
not point at are none the worse for its absence'.\footnote{This is not to
suggest that Brown and Wallace, or other proponents of the Claim, actually
make the Claim in the `strong' form given here. We consider this form first,
because it seems to us to be the strongest possible version of the argument.}
One might assert that here there really are two macroscopic worlds, one built
from $\Psi_{1}$ alone, and one built from $\Psi_{2}$ together with $q$. And
again, as in the case of the hydrogen atom discussed in section 5, one might
argue that there is no difference in the behaviour of these two worlds, and
that the motion of $\Psi_{1}$ represents a world every bit as \textit{bona
fide} as the world represented by $\Psi_{2}$ (together with $q$, which one
might assert is superfluous).

But again, as in the case of the hydrogen atom, pilot-wave theory tells us
that a remote experimenter with access to nonequilibrium particles could in
principle track the true history $q(t)$, without affecting $\Psi$. Further,
once it is known which packet is empty and which not, the experimenter could
perform additional experiments showing that $\Psi_{1}$ and $\Psi_{2}$
(predictably) behave \textit{differently} under certain operations. Again, the
empty packet is merely simulating a classical world, and the simulation holds
only under a class of operations more restrictive than those allowed in
pilot-wave theory. The situation is conceptually the same as in the case of
the single hydrogen atom.\footnote{Except, one might argue, if one is talking
about the `whole universe'. One could restrict the argument to
approximately-independent regions; this does not seem an essential point.}

We conclude that the Claim fails, even in a `strong form'.

\begin{center}
\textit{The Claim in a `Weak Form'}
\end{center}

Before considering more realistic approaches (with decoherence), it is
instructive to reconsider the above scenario in terms of a different --- and
equally unrealistic --- approach to the classical limit, namely the WKB
approach, in which the amplitude of $\Psi$ is taken to vary slowly over
relevant lengthscales. It is often said that the resulting wave function may
be `associated with' a family of classical trajectories, defined by the
equation $p=\nabla S$ giving the classical momentum $p$ in terms of the phase
gradient. (This approach is frequently used, for example, in quantum
cosmology.) Where such trajectories come from is not clear in standard quantum
theory, but in pilot-wave theory it is clear enough: in the WKB regime, the de
Broglie-Bohm trajectory $q(t)$ (within the extended wave) will indeed follow a
classical trajectory defined by $p=\nabla S$.

Now let the superposition%
\[
\Psi(q,t)=\frac{1}{\sqrt{2}}\left(  \Psi_{1}(q,t)+\Psi_{2}(q,t)\right)
\]
be composed of two non-overlapping `WKB packets' $\Psi_{1}$, $\Psi_{2}$, that
formed from the division of a single WKB packet $\Psi$, where again $q$
represents the contents of a macroscopic region including the Earth. As in the
earlier example, we imagine that the division occurred because a quantum
experiment was performed, with two possible outcomes indicated by a pointer
coordinate $y$: and again, $\Psi_{1}$ corresponds to the outcome $+1$, while
$\Psi_{2}$ corresponds to the outcome $-1$, and the actual $q(t)$ ends in the
support of $\Psi_{2}$. Unlike the earlier example, though, in this case the
packets $\Psi_{1}$, $\Psi_{2}$ are narrow with respect to $y$ but broad with
respect to the other (relevant) degrees of freedom --- so broad, in fact, that
with respect to these other degrees of freedom the packets are effectively
plane waves. The only really significant difference between $\Psi_{1}$ and
$\Psi_{2}$ is in their support with respect to $y$. (See Fig. 2.)%

\begin{figure}
[ptb]
\begin{center}
\includegraphics[
natheight=5.729400in,
natwidth=9.375400in,
height=2.041in,
width=3.3217in
]%
{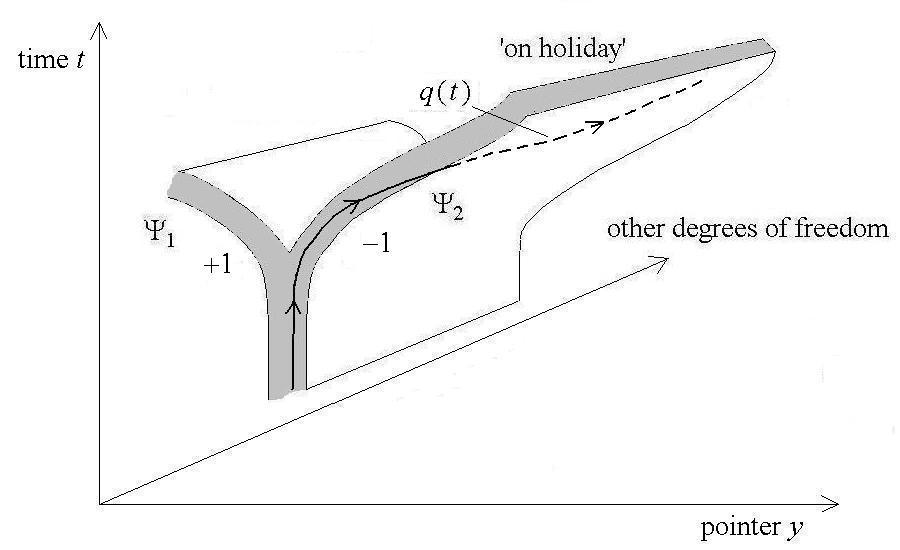}%
\caption{The Claim in a `weak form'.}%
\end{center}
\end{figure}

To be sure, this is not a realistic model of the macroscopic world, no more
than the Ehrenfest model was. But it is instructive to see the effect this
alternative approach has on the Claim.

Under the above assumptions, the actual trajectory $q(t)$ will be
approximately classical (except in the small branching region), and might be
taken to correctly model the macroscopic history with outcome $-1$ and the
experimenter going on holiday. But is there now any other discernible
realisation of an alternative classical macroscopic motion, such as the
experimenter staying at home? Clearly not. While the empty branch $\Psi_{1}$
is ontological, it is spread out over all degrees of freedom except $y$, so
that its time evolution does \textit{not} trace out a trajectory corresponding
to an approximately classical alternative motion. The experimenter `staying at
home' is nowhere to be seen. Unlike in the Ehrenfest case, one cannot point to
some piece of localised `$\Psi$-stuff' following an alternative classical trajectory.

Of course, different initial configurations $q(0)$ (with the same initial
$\Psi$) would yield different trajectories $q(t)$. And the `information' about
these alternative paths certainly exists in a mathematical sense, in the
structure of the complex field $\Psi$. But there is no reason to ascribe
anything other than mathematical status to these alternative trajectories ---
just as we saw in section 2, for the analogous classical case of a test
particle moving in an external electromagnetic field or in a background
spacetime geometry. The alternative trajectories are mathematical, not ontological.

\begin{center}
\textit{Realistic Models (with Environmental Decoherence)}
\end{center}

A more realistic account of the macroscopic, approximately classical realm may
be obtained from models with environmental decoherence. (For a review, see
Zurek (2003).)

Consider a system with configuration $q$, coupled to environmental degrees of
freedom $y=(x_{1},x_{2},...,x_{N})$. For a pure state the wave function is
$\Psi(q,y,t)$, and one often considers mixtures with a density operator%
\[
\hat{\rho}(t)=\sum_{\alpha}p_{\alpha}|\Psi_{\alpha}(t)\rangle\langle
\Psi_{\alpha}(t)|\ .
\]
(For example, in `quantum Brownian motion', the system is a single particle in
a potential and the environment consists of a large number of harmonic
oscillators in a thermal state.) By tracing over $y$ one obtains a reduced
density operator for the system, with matrix elements%
\[
\rho_{\mathrm{red}}(q,q^{\prime},t)\equiv\sum_{\alpha}p_{\alpha}\int
dy\ \Psi_{\alpha}(q,y,t)\Psi_{\alpha}^{\ast}(q^{\prime},y,t)\ ,
\]
from which one may define a quasi-probability distribution in phase space for
the system:%
\[
W_{\mathrm{red}}(q,p,t)\equiv\frac{1}{2\pi}\int dz\ e^{ipz}\rho_{\mathrm{red}%
}(q-z/2,q+z/2,t)
\]
(the reduced Wigner function). In certain conditions, one obtains an
approximately non-negative function $W_{\mathrm{red}}(q,p,t)$ whose time
evolution approximates that of a classical phase-space distribution.

For some elementary systems, such as a harmonic oscillator, the motion of a
narrowly-localised packet $W_{\mathrm{red}}(q,p,t)$ can trace out a thin
`tube' approximating a classical trajectory in phase space (Zurek \textit{et
al}. 1993). However, such simple quantum-classical correspondence breaks down
for chaotic systems, because of the rapid spreading of the packet: even an
initial minimum-uncertainty packet spreads over macroscopic regions of phase
space within experimentally-accessible timescales (Zurek 1998). On the other
hand, at least for some examples it can be shown that, even in the chaotic
case, the evolution of $W_{\mathrm{red}}(q,p,t)$ approximates the evolution of
a classical phase-space distribution $W_{\mathrm{class}}(q,p,t)$ (a Liouville
flow with a diffusive contribution from the environment), where both
distributions \textit{rapidly delocalise} (Habib \textit{et al}. 1998; Zurek
2003, pp. 745--47).\footnote{The examples are based on the weak-coupling,
high-temperature limit of quantum Brownian motion. The system consists of a
single particle moving in one dimension in a classically-chaotic potential.}

In pilot-wave theory, a mixed quantum state is described by a preferred
decomposition of $\hat{\rho}$ into a statistical mixture (with weights
$p_{\alpha}$) of ontological pilot waves $\Psi_{\alpha}$ (Bohm and Hiley
1996). For a given element of the ensemble, the de Broglian velocity of the
configuration is determined by the actual pilot wave $\Psi_{\alpha}$. (A
different decomposition generally yields different velocities, and so is
physically distinct at the fundamental level.) Now, the pilot-wave theory of
quantum Brownian motion has been studied by Appleby (1999). Under certain
conditions it was found that, as a result of decoherence, the de Broglie-Bohm
trajectories of the system become approximately classical (as one might have
expected). While Appleby made some simplifying assumptions in his analysis,
pending further studies of this kind it is reasonable to assume that Appleby's
conclusions hold more generally.

We may now evaluate the Claim in the context of realistic models. First of
all, as in the unrealistic examples considered above, the Claim fails because
an ideal subquantum measurement will always show that there is just one
trajectory $q(t)$; and, further experiments will show that empty wave packets
(predictably) behave differently from packets containing the actual
configuration. This alone suffices to refute the Claim. Even so, it is
interesting to ask if it is possible to have localised ontological packets
(`built out of $\Psi$') whose motions execute alternative classical histories:
that is, it is interesting to ask if the `strong form' of the Claim discussed
above --- which in any case fails, but is still rather intriguing --- could
ever occur in practice in realistic models. The answer, again, is no.

For an elementary non-chaotic system, one can obtain a narrow `Wigner packet'
$W_{\mathrm{red}}(q,p,t)$ approximating a classical trajectory, and one could
also have a superposition of two or more such packets (with macroscopic
separations). One might then argue that, since $W_{\mathrm{red}}$ is built out
of $\Psi$, we have (in a realistic setting, with decoherence) something like
the `strong form' of the Claim discussed above. However, the models usually
involve a mixture of $\Psi$'s, of which $W_{\mathrm{red}}$ is not a local
functional. So the ontological status of a narrow packet $W_{\mathrm{red}}$ is
far from clear. But even glossing over this, having a narrow packet
$W_{\mathrm{red}}$ following an approximately classical path is in any case
unrealistic in a world containing chaos, where, as we have already stated, one
can show only that $W_{\mathrm{red}}$ --- an approximately non-negative
function, with a large spread over phase space --- has a time evolution that
approximately agrees with the time evolution of a classical (delocalised)
phase-space distribution; that is, $W_{\mathrm{red}}$ follows an approximately
Hamiltonian or Liouville flow (with a diffusive contribution). Again, one
cannot obtain anything like `localised ontological $\Psi$-stuff' (or something
locally derived therefrom) executing an approximately classical trajectory ---
not even for one particle in a chaotic potential, and certainly not for a
realistic world containing turbulent fluid flow, double pendulums, people,
wars, and so on.

One \textit{can} obtain localised trajectories from a quantum description of a
chaotic system, if the system is continuously measured --- which in practice
involves an experimenter continuously monitoring an apparatus or environment
that is interacting with the system (Bhattacharya \textit{et al}. 2000). Such
trajectories for the Earth and its contents might in principle be obtained by
monitoring the environment (the interstellar medium, the cosmic microwave
background, etc.), but in the absence of an experimenter performing the
required measurements it is difficult to see how this could be relevant to our
discussion. And in any case, in a pilot-wave treatment, there is no reason why
such a procedure would yield `localised ontological $\Psi$-stuff' executing
the said trajectories.

In a realistic quantum-theoretical model, then, the outcome is a highly
delocalised distribution $W_{\mathrm{red}}(q,p,t)$ on phase space, obeying an
approximately Hamiltonian or Liouville evolution (with a diffusive
contribution). As in the unrealistic WKB example above, in pilot-wave theory
there will be one trajectory for each system. And, while different initial
conditions will yield different trajectories, there is no reason to ascribe
anything other than mathematical status to these alternatives --- just as in
the analogous classical case of a test particle moving in an external field or
background geometry. Once again, the alternative trajectories are
mathematical, not ontological.

Of course, given such a distribution $W_{\mathrm{red}}(q,p,t)$, \textit{if one
wishes} one may identify the flow with a set of trajectories representing
parallel (approximately classical) worlds, as in the decoherence-based
approach to many worlds of Saunders and Wallace. This is fair enough from a
many-worlds point of view. But if we start from pilot-wave theory understood
on its own terms, there is no motivation for doing so: such a step would
amount to a reification of mathematical structure (assigning reality to all
the trajectories associated with the velocity field at all points in phase
space). If one does so reify, one has constructed a different physical theory,
with a different ontology; one may do so if one wishes, but from a pilot-wave
perspective there is no special reason to take this step.

\begin{center}
\textit{Other Approaches to Decoherence}
\end{center}

Finally, decoherence and the emergence of the classical limit has also been
studied using the decoherent histories formulation of quantum
theory.\footnote{See, for example, Gell-Mann and Hartle (1993) and Halliwell
(1998), as well as the reviews in this volume.} In these treatments, there
will still be no discernible `localised ontological $\Psi$-stuff' following
alternative classical trajectories, for realistic models containing chaos.
Therefore, again, the `strong form' of the Claim (which in any case fails by
virtue of subquantum measurement) could never occur in practice.

\section{Further Remarks}

\begin{center}
\textit{Many de Broglie-Bohm Worlds?}
\end{center}

In the Saunders-Wallace approach to many worlds, one ascribes reality to the
full set of trajectories associated with the reduced Wigner function
$W_{\mathrm{red}}(q,p,t)$ in the classical limit (for some
appropriately-defined macrosystem with configuration $q$). This raises a
question. Why not \textit{also} ascribe reality to the full set of de
Broglie-Bohm trajectories outside the classical limit, for arbitrary (pure)
quantum states, resulting in a theory of `many de Broglie-Bohm
worlds'?\footnote{Such a theory has, in effect, been considered by Tipler
(2006).}

After all, just as $W_{\mathrm{red}}$ has a natural velocity field associated
with it (on phase space), so an arbitrary wave function $\Psi$ has a natural
velocity field associated with it (on configuration space) --- namely, de
Broglie's velocity field derived from the phase gradient $\nabla S$ (or more
generally, from the quantum current). In both cases, the velocity fields
generate a set of trajectories, and one may ascribe reality to them all if one
wishes. Why do so in the first case, but not in the second?

Furthermore, if the results due to Appleby (1999) (mentioned in section 6) for
quantum Brownian motion hold more generally, the parallel de Broglie-Bohm
trajectories will reduce to the parallel classical trajectories in an
appropriate limit; in which case, the theory of `many de Broglie-Bohm worlds'
will reproduce the Saunders-Wallace multiverse in the classical limit, and
will provide a simple and natural extension of it outside that limit --- that
is, one will have a notion of parallel worlds that is defined generally, even
at the microscopic level, and not just in the classical-decoherent
limit.\footnote{One need not think of this as `adding' trajectories to the
wave function; one could think of it as an alternative reading of physical
structure already existing in the `bare' wave function.}

However, since the de Broglie velocity field is single-valued, trajectories
$q(t)$ cannot cross. There can be no splitting or fusion of worlds. The above
`de Broglie-Bohm multiverse' then has the same kind of `trivial' structure
that would be obtained if one reified all the possible trajectories for a
classical test particle in an external field: the parallel worlds evolve
independently, side by side. Given such a theory, on the grounds of Ockham's
razor alone, there would be a conclusive case for taking only one of the
worlds as real.

On this point we remark that, in Deutsch's version of the Claim, if his word
`grooves' is interpreted as referring to the set of de Broglie-Bohm
trajectories, then the Claim amounts to asserting that pilot-wave theory
implies the de Broglie-Bohm multiverse. But again, because the parallel worlds
never branch or fuse, it would be natural to reduce the theory to a
single-world theory with only one trajectory.

A theory of many de Broglie-Bohm worlds, then, can only be a mere curiosity
--- a foil, perhaps, against which to test conventional Everettian ideas, but
not a serious candidate for a physical theory. On the other hand, it appears
to provide the basis for an argument against the Saunders-Wallace multiverse.
For as we have seen, it is natural to extend the Saunders-Wallace multiverse
to a deeper and more general de Broglie-Bohm multiverse.\footnote{It might be
claimed that, outside the nonrelativistic domain, such an extension is neither
simple nor natural. However, the (deterministic) pilot-wave theory of
high-energy physics has achieved a rather complete (if not necessarily final)
state of development --- for recent progress see Colin (2003), Colin and
Struyve (2007), Struyve (2008), Struyve and Westman (2007), and Valentini
(2008c).} And this, in turn, reduces naturally to a single-universe theory ---
that is, to standard de Broglie-Bohm theory. Thus, we have an argument that
begins by extending the Saunders-Wallace worlds to the microscopic level, and
ends by declaring only one of the resulting worlds to be real.

\begin{center}
\textit{Quantum Nonequilibrium and Many Worlds}
\end{center}

Since pilot-wave theory generally violates the Born rule, while conventional
many-worlds theory (apparently) does not, on this ground alone any attempt to
argue that the two theories are really the same must fail. Further, if such
violations were discovered,\footnote{See Valentini (2007, 2008a,b) for recent
discussions of possible experimental evidence.} then Everett's theory would be
disproved and that of de Broglie and Bohm vindicated.

On the other hand, it might be suggested that violations of the Born rule
could be incorporated into an Everett-type framework, by adopting the theory
of `many de Broglie-Bohm worlds' sketched above. Restricting ourselves for
simplicity to the pure case, if one assumes a nonequilibrium probability
measure $P_{0}\neq|\Psi_{0}|^{2}$ on the set of (parallel)\ initial
configurations $q(0)$, then for as long as relaxation to quantum equilibrium
has yet to occur completely, one will obtain a nonequilibrium set of parallel
trajectories $q(t)$, and one expects (in general) to find violations of the
Born rule within individual parallel worlds.\footnote{On the other hand,
quantum equilibrium for a multi-component closed system implies the Born rule
for measurements performed on subsystems (Valentini 1991a, D\"{u}rr \textit{et
al}. 1992).} If one accepts this, then observation of quantum nonequilibrium
would not suffice to disprove many worlds (though of course conventional
Everettian quantum theory \textit{would} be disproved). On the other hand,
however, as stated above it is natural to reduce the theory of many de
Broglie-Bohm worlds to a single-world theory, and this is equally true in the
nonequilibrium case. Therefore, the de Broglie-Bohm multiverse would not
provide a plausible refuge for the Everettian faced with nonequilibrium phenomena.

Even so, it might be worth exploring the theory of many de Broglie-Bohm worlds
with a nonequilibrium measure, in particular to highlight the assumptions made
in the Deutsch-Wallace derivation of the Born rule (Deutsch 1999, Wallace 2003a).

\begin{center}
\textit{On Arguments Concerning `Structure'}
\end{center}

One might argue that the mathematical structure in the quantum state that is
reified by many-worlds theorists plays such an explanatory and predictive role
that it should indeed be regarded as real. To quote Wallace (2003b, p. 93):

\begin{quote}
A tiger is any pattern which behaves as a tiger. .... the existence of a
pattern as a real thing depends on the usefulness --- in particular, the
explanatory power and predictive reliability --- of theories which admit that
pattern in their ontology.
\end{quote}

However, the behaviour of a system depends on the allowed set of experimental
operations. If one considers subquantum measurements, the patterns reified by
many-worlds theorists will cease to be explanatory and predictive. From a
pilot-wave perspective, then, such mathematical patterns are explanatory and
predictive only in the confines of quantum equilibrium: outside that limited
domain, subquantum measurement theory would provide a more explanatory and
predictive framework.

At best, it can only be argued that, if approximately classical experimenters
are confined to the quantum equilibrium state, so that they are unable to
perform subquantum measurements, then they will encounter a phenomenological
\textit{appearance} of many worlds --- just as they will encounter a
phenomenological appearance of locality, uncertainty, and of quantum physics generally.

\begin{center}
\textit{On Arguments Concerning Computation}
\end{center}

It might be argued that quantum computation provides evidence for the
existence of many worlds (Deutsch 1985, 1997). Deutsch asks `how' and `where'
the supposedly huge number of parallel computations are performed, and has
challenged those who doubt the existence of parallel universes to provide an
explanation for quantum-computational processes such as Shor's factorisation
algorithm (Deutsch 1997, p. 217).

However, while it often used to be asserted that the advantages of quantum
computation originated from quantum superposition, the matter has become
controversial. Some workers, such as Jozsa (1998) and Steane (2003), claim
that entanglement is the truly crucial feature. Further, the ability to find
periods seems to be the mechanism underlying Shor's algorithm, and this is
arguably more related to the `wave-like' aspect of quantum physics than it is
to superposition (Mermin 2007).

Leaving such controversies aside, we know in any case that, in quantum
equilibrium, pilot-wave theory yields the same predictions as ordinary quantum
theory, including for quantum algorithms. In an assessment of precisely how
pilot-wave theory provides an explanation for a specific quantum algorithm, it
should be borne in mind that: (a) the theory contains an ontological pilot
wave propagating in many-dimensional configuration space; (b) the theory is
nonlocal; and (c) with respect to quantum `measurements', the theory is
contextual. There is then ample scope for exploring the pilot-wave-theoretical
account of quantum-computational processes, if one wishes to do so, just as
there is for any other type of quantum process.

\section{Counter-Claim: A General Argument Against Many Worlds}

We have refuted the Claim, that pilot-wave theory is `many worlds in denial'.
Here, we put forward a Counter-Claim:

\begin{itemize}
\item Counter-Claim: \textit{The theory of many worlds is unlikely to be true,
because it is ultimately motivated by the puzzle of quantum superposition,
which arises from a belief in eigenvalue realism, which is in turn based
(ultimately) on the intrinsically unlikely assumption that quantum
measurements should be modelled on classical measurements.}
\end{itemize}

We saw in section 3 that quantum theorists call an experiment `a measurement
of $\omega$' only because it formally resembles what \textit{would have been}
a correct measurement of $\omega$ had the system been classical. Thus, the
system-apparatus interaction Hamiltonian is chosen by means of (for example)
the mapping%
\begin{equation}
H=a\omega p_{y}\longrightarrow\hat{H}=a\hat{\omega}\hat{p}_{y}\ , \label{Map}%
\end{equation}
so that quantum `measurements' are in effect modelled on classical
measurements. That this is a mistake is clear from a pilot-wave
perspective.\footnote{In the classical limit of pilot-wave theory, emergent
effective degrees of freedom have a purely mathematical correspondence with
linear operators acting on the wave function. Physicists trapped in quantum
equilibrium have made the mistake of taking this correspondence literally
(Valentini 1992, pp. 14--16, 19--29; 1996, pp. 50--51).} But the key point is
more general, and does not depend on pilot-wave theory. In fact, it was made
by Einstein in 1926 (see below).

\begin{center}
\textit{The Argument}
\end{center}

Everett's initial motivation for introducing many worlds was the puzzle of
quantum superposition, in particular the apparent transfer of superposition
from microscopic to macroscopic scales during a quantum measurement (Everett
1973, pp. 4--6). While our understanding of the theory today differs in many
respects from Everett's, it is highly doubtful that the theory would ever have
been proposed, were it not for the puzzle of quantum superposition.

Now, the puzzle of superposition stems from what we have called `eigenvalue
realism': the assignment of an ontological status to the eigenvalues of linear
operators acting on the wave function. For if an initial wave function%
\[
\psi_{0}(x)=\sum_{n}c_{n}\phi_{n}(x)
\]
is a superposition of different eigenfunctions $\phi_{n}(x)$ of $\hat{\omega}$
with different eigenvalues $\omega_{n}$, then if one takes eigenvalue realism
literally it appears as if all the values $\omega_{n}$ should somehow be
regarded as simultaneous ontological attributes of a single system.

Why do so many physicists believe in eigenvalue realism? The answer lies,
ultimately, in their belief in the quantum theory of measurement. For example,
it is widely thought that an experimental operation described by the
Hamiltonian operator $\hat{H}=a\hat{\omega}\hat{p}_{y}$ constitutes a correct
measurement of an observable $\omega$, as indicated by the value of the
pointer coordinate $y$. To see that this leads to a belief in eigenvalue
realism, consider a system with wave function $\phi_{n}(x)$. Under such an
operation, the pointer $y$ will indicate the value $\omega_{n}$. Because the
experimenter \textit{believes} that this pointer reading provides a correct
measurement, the experimenter will then believe that the system must have a
property $\omega$ with value $\omega_{n}$ --- that is, the experimenter will
believe in eigenvalue realism.

Now, why do so many physicists believe that an operation described by (for
example) $\hat{H}=a\hat{\omega}\hat{p}_{y}$ constitutes a correct measurement
of $\omega$, for any observable $\omega$? The answer, as we have seen, is that
the said operation formally resembles a classical measurement of $\omega$, via
the mapping (\ref{Map}).

We claim that this is the heart of the matter: it is widely assumed, in
effect, that classical physics provides a reliable guide to measurement for
nonclassical systems. We claim further that this assumption is intrinsically
unlikely, so that the conclusions stemming from it --- eigenvalue realism,
superposition of properties, multiplicity of worlds --- are in turn
intrinsically unlikely (Valentini 1992, pp. 14--16, 19--29; 1996, pp. 50--51).

The assumption is unlikely because, generally speaking, one cannot use a
theory as an accurate guide to measurement outside the domain of validity of
the theory. For experiment is theory-laden, and correct measurement procedures
must be laden with the correct theory. As an example, consider what might
happen if one used Newton's theory of gravity to interpret observations close
to a black hole: one would encounter numerous puzzles and paradoxes, that
would be resolved only when the observations were interpreted using general
relativity. It is intrinsically improbable that measurement operations taken
from an older, superseded physics will remain valid in a fundamentally new
domain for all possible observables. It is much more likely that a new domain
will be better understood in terms of a new theory based on new concepts, with
its own new theory of measurement --- as shown by the example of general
relativity, and indeed by the example of de Broglie's nonclassical
dynamics.\footnote{In contrast with Bohr's unwarranted claim: `The unambiguous
interpretation of any measurement must be essentially framed in terms of the
classical physical theories, and we may say that in this sense the language of
Newton and Maxwell will remain the language of physicists for all time' (Bohr
1931).}

\begin{center}
`\textit{Einstein's Hot Water'}
\end{center}

This very point was made by Einstein in 1926, in a well-known conversation
with Heisenberg (Heisenberg 1971, pp. 62--69). This conversation is often
cited as evidence of Einstein's view that observation is theory-laden. But a
crucial element is usually missed: Einstein also warned Heisenberg that his
treatment of observation was unduly laden with the superseded theory of
classical physics, and that this would eventually cause trouble (Valentini
1992, p. 15; 1996, p. 51).

During the conversation, Heisenberg made the (at the time fashionable) claim
that `a good theory must be based on directly observable magnitudes' (p. 63).
Einstein replied that, on the contrary (p. 63):

\begin{quote}
.... it is quite wrong to try founding a theory on observable magnitudes
alone. In reality the very opposite happens. \textit{It is the theory which
decides what we can observe}. [Italics added.]
\end{quote}
Einstein added that there is a long, complicated path underlying any
observation, which runs from the phenomenon, to the production of events in
our apparatus, and from there to the production of sense impressions. And
theory is required to make sense of this process:

\begin{quote}
Along this whole path .... we must be able to tell how nature functions ....
before we can claim to have observed anything at all. Only theory, that is,
knowledge of natural laws, enables us to deduce the underlying phenomena from
our sense impressions.
\end{quote}
Einstein's key point so far is that, as we have said, there is no
\textit{a priori} notion of how to perform a correct measurement: one requires
some knowledge of physics to do so. If we wish to design a piece of apparatus
that will correctly measure some property $\omega$ of a system, then we need
to know the correct laws governing the interaction between the system and the
apparatus, to ensure that the apparatus pointer will finish up pointing to the
correct reading. (One cannot, for example, design an ammeter to measure
electric current without some knowledge of electromagnetic forces.)

Now, Einstein went on to note that, when new experimental phenomena are
discovered --- phenomena that require the formulation of a new theory --- in
practice the old theory is at first assumed to provide a reliable guide to
interpreting the observations (pp. 63--64):

\begin{quote}
When we claim that we can observe something new, we ought really to be saying
that, although we are about to formulate new natural laws that do not agree
with the old ones, we nevertheless assume that the existing laws --- covering
the whole path from the phenomenon to our consciousness --- function in such a
way that we can rely upon them and hence speak of `observations'.
\end{quote}
Note that this is a practical necessity, for the new theory has yet to
be formulated. However --- and here is the crucial point --- once the new
theory \textit{has} been formulated, one ought to be careful to use the new
theory to design and interpret measurements, and not continue to rely on the
old theory to do so. For one may well find that consistency is obtained only
when the new laws are found \textit{and applied to the process of
observation}. If one fails to do this, one is likely to cause difficulties.
That Einstein saw this very point is clear from a subsequent passage (p. 66):

\begin{quote}
I have a strong suspicion that, precisely because of the problems we have just
been discussing, your theory will one day get you into hot water. .... When it
comes to observation, you behave as if everything can be left as it was, that
is, as if you could use the old descriptive language.
\end{quote}
Here, then, is Einstein's warning to Heisenberg: not to interpret
observations of quantum systems using the `old descriptive language' of
classical physics. The point, again, is that while observation is in general
theory-laden, in quantum theory observations are incorrectly laden with a
\textit{superseded }theory (classical physics), and this will surely lead to trouble.

We claim that the theory of many worlds is precisely an example of what one
might call `Einstein's hot water'. Specifically, the apparent multiplicity of
the quantum domain is an illusion, caused by an over-reliance on a superseded
(classical) physics as a guide to observation and measurement --- a mistake
that is the ultimate basis of the belief in eigenvalue realism, which in turn
led to the puzzle of superposition and to Everett's valiant attempt to resolve
that puzzle.

\section{Conclusion}

Pilot-wave theory is intrinsically nonclassical, with its own (`subquantum')
theory of measurement, and it is in general a `nonequilibrium' theory that
violates the quantum Born rule. From the point of view of pilot-wave theory
itself, an apparent multiplicity of worlds at the microscopic level (envisaged
by some theorists) stems from the generally mistaken assumption that
eigenvalues have an ontological status (`eigenvalue realism'), which in turn
ultimately derives from the generally mistaken assumption that `quantum
measurements' are true and proper measurements.

At the macroscopic level, it might be thought that the universal (and
ontological) pilot wave can develop non-overlapping and localised branches
that evolve just like parallel classical worlds. But in fact, such localised
branches are unrealistic (especially over long periods of time, and even for
short periods of time in a world containing chaos). And in any case,
subquantum measurements could track the actual de Broglie-Bohm trajectory, so
that in principle one could distinguish the branch containing the
configuration from the empty ones, where the latter would be regarded merely
as concentrations of a complex-valued configuration-space field.

In realistic models of decoherence, the pilot wave is delocalised, and the
identification of a set of parallel (approximately) classical worlds does not
arise in terms of localised pieces of actual `$\Psi$-stuff' executing
approximately classical motions. Instead, such identification amounts to a
reification of purely mathematical trajectories --- a move that is fair enough
from a many-worlds perspective, but which is unnecessary and unjustified from
a pilot-wave perspective because according to pilot-wave theory there is
nothing actually moving along any of the trajectories except one (just as in
the classical theory of a test particle in an external field or background
spacetime geometry). In addition to being unmotivated, such reification begs
the question of why the mathematical trajectories should not also be reified
outside the classical limit for general wave functions, resulting in a theory
of `many de Broglie-Bohm worlds' (which in turn naturally reduces to a
single-world theory).

Properly understood, pilot-wave theory is not `many worlds in denial': it is a
different physical theory. Furthermore, from the perspective of pilot-wave
theory itself, many worlds are an illusion. And indeed, even leaving
pilot-wave theory aside, we have seen that the theory of many worlds is rooted
in the intrinsically unlikely assumption that quantum measurements should be
modelled on classical measurements, and is therefore in any case unlikely to
be true.

\textbf{Acknowledgements.} This work was partly supported by grant RFP1-06-13A
from the Foundational Questions Institute (fqxi.org). For their hospitality, I
am grateful to Carlo Rovelli and Marc Knecht at the Centre de Physique
Th\'{e}orique (Luminy), to Susan and Steffen Kyhl in Cassis, and to Jonathan
Halliwell at Imperial College London.

\begin{center}
BIBLIOGRAPHY
\end{center}

Appleby, D. M. (1999). Bohmian trajectories post-decoherence.
\textit{Foundations of Physics}, \textbf{29}, 1885--1916.

Bacciagaluppi, G. and Valentini, A. (2009). \textit{Quantum Theory at the
Crossroads: Reconsidering the 1927 Solvay Conference}. Cambridge: Cambridge
University Press [quant-ph/0609184].

Bell, J. S. (1987). \textit{Speakable and Unspeakable in Quantum Mechanics}.
Cambridge: Cambridge University Press.

Bhattacharya, T., Habib, S., and Jacobs, K. (2000). Continuous quantum
measurement and the emergence of classical chaos. \textit{Physical Review
Letters}, \textbf{85}, 4852--4855.

Bohm, D. (1952a). A suggested interpretation of the quantum theory in terms of
`hidden' variables, I. \textit{Physical Review}, \textbf{85}, 166--179.

Bohm, D. (1952b). A suggested interpretation of the quantum theory in terms of
`hidden' variables, II. \textit{Physical Review}, \textbf{85}, 180--193.

Bohm, D. and Hiley, B. J. (1996). Statistical mechanics and the ontological
interpretation. \textit{Foundations of Physics}, \textbf{26}, 823--846.

Bohr, N. (1931). Maxwell and modern theoretical physics. \textit{Nature},
\textbf{128}, 691--692. Reprinted in \textit{Niels Bohr: Collected Works},
vol. 6, ed. J. Kalckar. Amsterdam: North-Holland, 1985, p. 357.

Brown, H. R. and Wallace, D. (2005). Solving the measurement problem: de
Broglie--Bohm loses out to Everett. \textit{Foundations of Physics},
\textbf{35}, 517--540.

Colin, S. (2003). A deterministic Bell model. \textit{Physics Letters A},
\textbf{317}, 349--358 [quant-ph/0310055].

Colin, S. and Struyve, W. (2007). A Dirac sea pilot-wave model for quantum
field theory. \textit{Journal of Physics A: Mathematical and Theoretical},
\textbf{40}, 7309--7341 [quant-ph/0701085].

de Broglie, L. (1928). La nouvelle dynamique des quanta. In
\textit{\'{E}lectrons et Photons: Rapports et Discussions du Cinqui\`{e}me
Conseil de Physique}. Paris: Gauthier-Villars, pp. 105--132. [English
translation: Bacciagaluppi, G. and Valentini, A. (2009).]

Deutsch, D. (1985). Quantum theory, the Church-Turing principle and the
universal quantum computer. \textit{Proceedings of the Royal Society of London
A}, \textbf{400}, 97--117.

Deutsch, D. (1986). Interview. In \textit{The Ghost in the Atom}, eds. P. C.
W. Davies and J. R. Brown. Cambridge: Cambridge University Press, pp. 83--105.

Deutsch, D. (1996). Comment on Lockwood. \textit{British Journal for the
Philosophy of Science}, \textbf{47}, 222--228.

Deutsch, D. (1997). \textit{The Fabric of Reality}. London: Penguin.

Deutsch, D. (1999). Quantum theory of probability and decisions.
\textit{Proceedings of the Royal Society of London A}, \textbf{455}, 3129--3137.

D\"{u}rr, D., Goldstein, S., and Zangh\`{\i}, N. (1992). Quantum equilibrium
and the origin of absolute uncertainty. \textit{Journal of Statistical
Physics}, \textbf{67}, 843--907.

D\"{u}rr, D., Goldstein, S., and Zangh\`{\i}, N. (1996). Bohmian mechanics as
the foundation of quantum mechanics. In \textit{Bohmian Mechanics and Quantum
Theory: an Appraisal}, eds. J. T. Cushing \textit{et al}.. Dordrecht: Kluwer,
pp. 21--44.

D\"{u}rr, D., Goldstein, S., and Zangh\`{\i}, N. (1997). Bohmian mechanics and
the meaning of the wave function. In \textit{Experimental Metaphysics: Quantum
Mechanical Studies for Abner Shimony}, eds. R. S. Cohen \textit{et al}..
Dordrecht: Kluwer, pp. 25--38.

Everett, H. (1973). The theory of the universal wave function. In \textit{The
Many-Worlds Interpretation of Quantum Mechanics}, eds. B. S. DeWitt and N.
Graham. Princeton: Princeton University Press, pp. 3--140.

Gell-Mann, M., and Hartle, J. B. (1993). Classical equations for quantum
systems. \textit{Physical Review D}, \textbf{47}, 3345--3382.

Habib, S., Shizume, K., and Zurek, W. H. (1998). Decoherence, chaos, and the
correspondence principle. \textit{Physical Review Letters}, \textbf{80}, 4361--4365.

Halliwell, J. J. (1998). Decoherent histories and hydrodynamic equations.
\textit{Physical Review D}, \textbf{58}, 105015.

Heisenberg, W. (1971). \textit{Physics and Beyond}. New York: Harper {\&} Row.

Holland, P. R. (1993). \textit{The Quantum Theory of Motion: An Account of the
de Broglie-Bohm Causal Interpretation of Quantum Mechanics}. Cambridge:
Cambridge University Press.

Jozsa, R. (1998). Entanglement and quantum computation. In \textit{The
Geometric Universe: Science, Geometry, and the Work of Roger Penrose}. Oxford:
Oxford University Press, pp. 369--379.

Mermin, N. D. (2007). What has quantum mechanics to do with factoring?
\textit{Physics Today}, \textbf{60} (April 2007), 8--9.

Pearle, P., and Valentini, A. (2006). Quantum mechanics: generalizations. In
\textit{Encyclopaedia of Mathematical Physics}, eds. J.-P. Fran\c{c}oise
\textit{et al}.. Amsterdam: Elsevier, pp. 265--76 [quant-ph/0506115].

Rovelli, C. (2004). \textit{Quantum Gravity}. Cambridge: Cambridge University Press.

Steane, A. M. (2003). A quantum computer only needs one universe.
ArXiv:quant-ph/0003084v3 (24 March 2003).

Struyve, W. (2008). De Broglie-Bohm field beables for quantum field theory.
\textit{Physics Reports} (to appear) [arXiv:0707.3685].

Struyve, W. and Valentini, A. (2008). De Broglie-Bohm Guidance Equations for
Arbitrary Hamiltonians. \textit{Journal of Physics A: Mathematical and
Theoretical} (to appear) [arXiv:0808.0290].

Struyve, W. and Westman, H. (2007). A minimalist pilot-wave model for quantum
electrodynamics. \textit{Proceedings of the Royal Society of London A},
\textbf{463}, 3115--3129 [arXiv:0707.3487].

Tipler, F. J. (1987). Non-Schr\"{o}dinger forces and pilot waves in quantum
cosmology. \textit{Classical and Quantum Gravity}, \textbf{4}, L189--L195.

Tipler, F. J. (2006). What about quantum theory? Bayes and the Born
interpretation. ArXiv:quant-ph/0611245.

Valentini, A. (1991a). Signal-locality, uncertainty, and the subquantum
\textit{H}-theorem, I. \textit{Physics Letters A}, \textbf{156}, 5--11.

Valentini, A. (1991b). Signal-locality, uncertainty, and the subquantum
\textit{H}-theorem, II. \textit{Physics Letters A}, \textbf{158}, 1--8.

Valentini, A. (1992). On the pilot-wave theory of classical, quantum and
subquantum physics. Ph.D. thesis, International School for Advanced Studies,
Trieste, Italy [http://www.sissa.it/ap/PhD/Theses/valentini.pdf].

Valentini, A. (1996). Pilot-wave theory of fields, gravitation and cosmology.
In \textit{Bohmian Mechanics and Quantum Theory: an Appraisal}, eds. J. T.
Cushing \textit{et al}.. Dordrecht: Kluwer, pp. 45--66.

Valentini, A. (2001). Hidden variables, statistical mechanics and the early
universe. In \textit{Chance in Physics: Foundations and Perspectives}, eds. J.
Bricmont \textit{et al}.. Berlin: Springer-Verlag, pp. 165--81 [quant-ph/0104067].

Valentini, A. (2002b). Subquantum information and computation. \textit{Pramana
--- Journal of Physics}, \textbf{59}, 269--277 [quant-ph/0203049].

Valentini, A. (2004a). Universal signature of non-quantum systems.
\textit{Physics Letters A}, \textbf{332}, 187--193 [quant-ph/0309107].

Valentini, A. (2004b). Black holes, information loss, and hidden variables. ArXiv:hep-th/0407032.

Valentini, A. (2007). Astrophysical and cosmological tests of quantum theory.
\textit{Journal of Physics A: Mathematical and Theoretical}, \textbf{40},
3285--3303 [hep-th/0610032].

Valentini, A. (2008a). Inflationary cosmology as a probe of primordial quantum
mechanics. ArXiv:0805.0163.

Valentini, A. (2008b). De Broglie-Bohm prediction of quantum violations for
cosmological super-Hubble modes. ArXiv:0804.4656.

Valentini, A. (2008c). \textit{International Journal of Modern Physics A} (to appear).

Valentini, A. and Westman, H. (2005). Dynamical origin of quantum
probabilities. \textit{Proceedings of the Royal Society of London A},
\textbf{461}, 253--272.

Wallace, D. (2003a). Everettian rationality: defending Deutsch's approach to
probability in the Everett interpretation. \textit{Studies in History and
Philosophy of Modern Physics}, \textbf{34}, 415--439.

Wallace, D. (2003b). Everett and structure. \textit{Studies in History and
Philosophy of Modern Physics}, \textbf{34}, 87--105.

Zeh, H. D. (1999). Why Bohm's Quantum Theory? \textit{Foundations of Physics
Letters}, \textbf{12}, 197--200.

Zurek, W. H. (1998). Decoherence, chaos, quantum-classical correspondence, and
the algorithmic arrow of time. \textit{Physica Scripta}, \textbf{T76}, 186--198.

Zurek, W. H. (2003). Decoherence, einselection, and the quantum origins of the
classical. \textit{Reviews of Modern Physics}, \textbf{75}, 715--775.

Zurek, W. H., Habib, S., and Paz, J. P. (1993). Coherent states via
decoherence. \textit{Physical Review Letters}, \textbf{70}, 1187--1190.

\end{document}